\documentclass[reprint,floatfix,aps,prd,showpacs,footinbib,amsmath,amssymb,amsfonts,superscriptaddress]{revtex4-1}
\usepackage{bm}
\usepackage{graphicx}
\usepackage{color}
\usepackage{dsfont}
\usepackage[normalem]{ulem}
\usepackage{hyperref}
\usepackage{mathtools}
\usepackage{mathrsfs}
\usepackage{calrsfs}
\usepackage{comment}
\usepackage[utf8]{inputenc}
\usepackage{varioref}
\usepackage[active]{srcltx}

\expandafter\ifx\csname package@font\endcsname\relax\else
\expandafter\expandafter
\expandafter\usepackage
\expandafter\expandafter
\expandafter{\csname package@font\endcsname}%
\fi
\hypersetup{
	colorlinks=true,       
	linkcolor=blue,          
	citecolor=blue,        
}

\usepackage[section]{placeins}

\usepackage{fancybox}
\labelformat{figure}{Fig.~#1} 

\begin{document}
	\title{Distinguishing general relativity from Chern-Simons gravity using 
gravitational wave polarizations} 
	\author{Soham Bhattacharyya} 
	\affiliation{School of Physics, Indian Institute of Science Education and Research Thiruvananthapuram (IISER-TVM), Trivandrum 695551, India}
	\email{xeonese13@iisertvm.ac.in}
	\author{S. Shankaranarayanan}
	\affiliation{Department of Physics, Indian Institute of Technology Bombay, Mumbai 400076, India}
	\email{shanki@phy.iitb.ac.in}
\begin{abstract}
Quasi-normal modes (QNMs) uniquely characterize the final black-hole. Till now, only the QNM frequency and damping time are used to test General relativity. In this work, we show explicitly that another property of the QNMs --- \emph{their polarization} ---  can be a reliable tool for probing gravity. We provide a consistent test for General relativity by considering Chern-Simons gravity. Distinguishing Chern-Simons gravity from General relativity using only template matching is highly challenging. Thus a parameter that can differentiate between Chern-Simons gravity and GR will be a suitable candidate for any modified theories of gravity. We discuss the implications of our result for the future gravitational wave detectors. 
\end{abstract}
	\pacs{}
	\maketitle

\section{Introduction}
Direct detections of gravitational waves (GW) from compact binary objects have energized searches for deviations from the general theory of relativity (GR)~\cite{Abbott2016a,Berti:2015itd,Barack:2018yly}. {In Ref. \cite{PhysRevLett.116.221101},  constraints on the graviton mass were obtained based on dispersion in a vacuum, and a more general prescription was given in \cite{Cardoso:2018zhm}}. These constraints were obtained by matching templates from numerical simulations of GR with observed data, and introducing new parameters corresponding to extended gravity theories~\cite{Sathyaprakash:2009xs,Gair:2012nm,Yunes:2013dva}. However, as the accuracy of the current and upcoming detectors (including LISA) increase, the sensitivities of template matching techniques will saturate, and there is an urgent need to find alternative strategies to test for deviations from GR.

Even if waveforms for the numerous modifications to gravity~\cite{Achour2018,Clifton2012,Nojiri:2006ri} can be obtained, it is imperative to obtain a handful of parameters that can be used as a consistency test of GR. Specifically, it is essential to find a dimensionless, model-independent parameter which vanishes for GR and finite for modified gravity theories. Such parameters have been constructed to distinguish between dark energy models and modified gravity theories~(see, for instance, \cite{Song:2008vm}).
    
{Gravitational waves emitted by perturbed black holes during the ring-down epoch are mathematically described by Quasi-normal modes (QNMs), and are fingerprints of the final black-hole as they depend \emph{only} on parameters characterizing the BH (like mass, charge and angular momentum)}~\cite{Berti2009,Konoplya2011,Nollert:1999ji,Kokkotas1999}. Thus, extracting the frequencies and damping times allow one to test GR \cite{Berti:2015itd,Yunes:2016jcc,Baibhav:2017jhs}. However, another property of the QNMs, their polarizations, can be a reliable tool for probing gravity. Recently, the current authors used an inequality between polar and axial gravitational perturbations in $f(R)$ theories to obtain a parameter vanishing for GR and finite for $f(R)$ theories~\cite{Bhattacharyya2017,Bhattacharyya2018}. In this study, we propose a parameter to distinguish GR from (dynamical) Chern Simons (CS) Gravity and show that the inequality between polar and axial perturbations is, \emph{model-independent} and, valid for any modification to GR.

CS Gravity is indistinguishable from GR for all conformally flat space-times and space-times that possess a maximally symmetric 2-dimensional subspace \cite{Campbell:1990fu}. Thus a parameter that can distinguish between CS gravity and GR will be a suitable candidate for any modified theories of gravity. 
Naturally, of late, there is a lot of interest in studying the perturbations about Schwarzschild and slowly-rotating black-holes in dynamical CS gravity~\cite{Smith:2007jm,Molina2010,Cardoso2009,Motohashi2011,Kimura:2018nxk,Okounkova:2017yby} and, more recently, in Ref. \cite{Okounkova:2018pql}.

In this article, we show that isospectrality \cite{Chandrasekhar:1985kt} between odd and even parity perturbations is broken for a perturbed Schwarzschild black-hole, and slowly rotating, in dynamical Chern-Simons (dCS) gravity in a gauge invariant manner~\cite{Kodama:2000fa}. Consequently, odd-even parities carry different amounts of gravitational energy to asymptotic infinity. We quantify the relative difference between the two by constructing an energy-momentum pseudotensor of perturbation and show that the modification to gravitational radiation is more significant around the black-hole region compared to flat space-times. 
Also, we remark that the quantifying parameter can distinguish between GR and any modified theory of gravity.

For easy comparison, we use the notations/conventions of Ref.~\cite{Martel2005}.
We use $ \left(-,+,+,+\right) $ signature, Greek for the 4-D space-time, upper case Latin for the $ \left(\theta,\phi\right) $ and lower case Latin for $ \left(t,r\right) $, c = G = 1 such that $ \kappa^2 = 8\pi $. $ \nabla $ and subscript semicolon are covariant derivatives of the full space-time, $ D $ is the covariant derivative for $ \left(t,r\right) $, $ \hat{D} $ is the covariant derivative on a 2-sphere, and $ \Omega\equiv\Omega(\theta,\phi) $ denote the coordinates on a 2-sphere. Overbarred quantities are background and $ A^{(i)} $ denotes the $i^{th} $ order perturbation of the object $A$.
    
\section{Perturbations in GR}
We consider Schwarzschild space-time as our background $ \bar{g}_{\mu\nu} $.
    \begin{eqnarray}
    ds^2 &=& \bar{g}_{ab}dx^adx^b + \bar{g}_{AB} dz^Adz^B \\
    &=& -f(r)dt^2 + \frac{dr^2}{f(r)} + r^2 d\Omega^2 \label{scbg}\\
    f(r) \equiv f &=& 1-\frac{2M}{r}
    \end{eqnarray}
    Here, the background space-time is split into a $ (t,r) $ space and a 2-sphere $ (\theta,\phi) $. The metric perturbations ($h_{\mu\nu}$)
    \begin{eqnarray}
    g_{\mu\nu} = \bar{g}_{\mu\nu} + \epsilon h_{\mu\nu} ;\qquad 
    g^{\mu\nu} = \bar{g}^{\mu\nu} - \epsilon h^{\mu\nu}
    \end{eqnarray}
    can be separated using spherical harmonics, {while $\epsilon $ is a smallness factor which ensure the effect of $ h_{\mu\nu} $ remains small and does not substantailly change the background}. The spherical harmonic functions corresponding to scalar, vector, and tensor components of $ h_{\mu\nu} $ are of two opposite parities, odd and even. Gravitational or scalar field perturbations thus reduce to a one dimensional scattering problem of the form \cite{Regge:1957td,Zerilli1970a,Chandrasekhar:1985kt,Kodama2003,Kodama2004b,Martel2005}
    \begin{eqnarray}
    &&\frac{d^2\Phi_i}{dr_*^2} + \left(\omega^2 - V_i\right)\Phi_i = 0 \label{grptb}\\
    &&i=\text{scalar, odd/even gravitational} \, , \nonumber
    \end{eqnarray}
    where $ r_* $ is the tortoise coordinate, and $ V_i $ are effective potentials induced by the background space-time curvature, depending on the type of perturbation (gravitational or scalar). {Dynamics of space-time around a ringing black hole can be replaced by a \emph{wave scattering off of a central potential} problem. For vacuum space-times (like post merger of a binary black hole system) the individual profiles of $ V_i $ determine the fraction of incident gravitational radiation that escapes to infinity.} For gravitational perturbations in GR, the profiles of $ V_i $ for both odd and even parity perturbations are same owing to an isospectral relationship that exists between them \cite{Chandrasekhar:1985kt}. This, along with the fact that the dynamics of the two parities remain decoupled at the linear order, leads to the conclusion that {the ratio of scattered/radiated gravitational energies through the two opposite parities will be a constant throughout the duration of the ring-down} \cite{Bhattacharyya2017} --- a feature that has important consequences for non-Einsteinian theories of gravity.

\section{Perturbations in Dynamical Chern-Simons}
The lowest order parity violating coupling term determined by a dynamical scalar field will have an action of the form~\cite{1982AnPhy.140..372D,Deser:1982vy,Jackiw2003,Smith:2007jm}. 
    \begin{eqnarray}
        S &=& \int d^4x \, \sqrt{-g} \left[\frac{R}{2\kappa^2} + \frac{\alpha}{4}\vartheta{}^*RR - \frac{\beta}{2}\left(\nabla\vartheta\right)^2 - \frac{\beta}{2}V\left(\vartheta\right)\right] \nonumber\\\label{action}
    \end{eqnarray}
    where $ \vartheta $ is a dynamical pseudo-scalar field. We have chosen $ \vartheta $ to be dimensionless, which leads to $ \left[\alpha\right] = L^2 $ and $ \left[\beta\right] $ is dimensionless and $ {}^*RR $ is 
    \begin{eqnarray}
        {}^*RR &=& \frac{1}{2} R_{\mu\nu\rho\sigma}\epsilon^{\mu\nu\alpha\beta}R^{\rho\sigma}_{\quad\alpha\beta}
    \end{eqnarray}
    referred to as Pontryagin density quantifying the extent to which local Lorentz invariance is violated. 
    For spherically symmetric space-times, the Pontryagin density vanishes, making the Schwarzschild space-time a solution of the CS modified field equations. Only recently, non-slow rotation black-hole space-times have been constructed~\cite{Delsate:2018ome}. For slow-rotating perturbative solutions, see \cite{Yunes2009,Yagi:2012ya}. In literature, one sets $\vartheta = V(\vartheta) = 0 $~\cite{Molina2010,Cardoso2009}. 

Expanding the metric perturbation and the pseudo-scalar using spherical harmonics~\cite{Martel2005}, two coupled equations characterize odd parity and CS field perturbations, while the even parity remains the same as in GR. The odd parity sector becomes:
    \begin{eqnarray}
    &&\frac{d^2\Phi_O}{dr_*^2} + \left(\omega^2 - V_O\right)\Phi_O = S^{eff} \label{eom1}\\
    &&\frac{d^2\varphi}{dr_*^2} + \left(\omega^2 - V_\varphi\right)\varphi = \frac{6\alpha \mu Mf}{\beta r^5}\Phi_O \label{eom2} \\
    &&\mu = \left(\ell-1\right)\ell\left(\ell+1\right)\left(\ell+2\right)
    \end{eqnarray}
$ \varphi $ is related to $ \vartheta $ as \cite{Cardoso2009,Molina2010}
    \begin{eqnarray}
    \vartheta\left(t,r,\Omega\right) = \frac{\varphi\left(r\right)}{r} \textbf{S}(\Omega) e^{i\omega t}  \label{vartovar}
    \end{eqnarray}
    and $ \textbf{S}(\Omega) $ is a scalar spherical harmonic function. { $ V_O $ and $ V_\varphi $ are the odd parity effective potential and the effective potential for a massless spin-0 field respectively. $ S^{eff} $ is of the form $ \sum_{0}^{2} a_n/r \partial^{n}_{r_*} \varphi $ and the functions $ a_n $ have been given in Appendix \ref{ptbcotton}. At asymptotic infinity the source terms of (\ref{eom1}) and (\ref{eom2}) vanish, and the two fields decouple.}
  
{Previously in \cite{Cardoso2009}, the preferential coupling of the CS scalar with the odd parity mode was found using the Regge-Wheeler gauge choice, however }to our knowledge, this is the first time a gauge-invariant analysis of black hole perturbations is done in CS gravity. {As was found previously by \cite{Cardoso2009}, isospectral relations break between the even and odd parity modes}. The reason for this breaking is the appearance of an inhomogeneous term in the RHS of (\ref{eom1}) and an absence of any such term in the even parity sector. Thus, the QNM frequencies of the even and odd parity modes will be different, {a feature that can be used as a test for deviations from GR. Furthermore, since the odd parity now couples with the CS field, it will exchange energy with the field, reducing the radiated energy through the odd parity mode compared to GR, while the radiated energy through the even parity mode remains the same as in GR. This imbalance is another feature that can be used to test for modifications to GR at strong gravity regimes.}

\section{Difference in energy flux}
{The ratio of radiated energies through odd and even parities will be different in CS theories compared to GR. The extent of the difference in ratios can be quantified by calculating the effective energy-momentum pseudotensor of perturbation in the curved background (for earlier work, see \cite{Stein2011a}).} We use Isaacson's shortwave approximation~\cite{Isaacson1968b,Isaacson1968a}. In this scheme, we average over the rapidly fluctuating spatial components of the metric perturbation compared to the length scales over which the background significantly changes and obtain the \emph{back-reaction} effect on the background metric $\bar{g}_{\mu\nu}$.
    
    The back-reaction effect on the background metric is given by
    \begin{eqnarray}
    \bar{\mathfrak{G}}_{\mu\nu} = - t_{\mu\nu} &\equiv& -2 \epsilon^2\kappa^2\alpha\left<C^{(2)}_{\mu\nu}\right> 
    - \epsilon^2\left<G^{(2)}_{\mu\nu}\right>  \nonumber\\
   & &  + \epsilon^2\kappa^2\beta \left<\vartheta_{;\mu}\vartheta_{;\nu}\right>
    \label{eq:Isaacsonapprox}
    \end{eqnarray}
    where $ G_{\mu\nu} $ is the Einstein tensor. It is useful to obtain $t_{\mu\nu}$ in the TT gauge and in-terms of the traced-reversed perturbed tensor $\psi_{\mu\nu} = h_{\mu\nu} - \frac{h}{2}\bar{g}_{\mu\nu}$, where $ h $ is the trace of $ h_{\mu\nu} $. The first order perturbed field equations are
    \begin{small}
        \begin{eqnarray}
        &&\Box \psi_{\mu\nu} + 2 \bar{R}_{\mu\alpha\nu\beta}\psi^{\alpha\beta} = 2\kappa^2\alpha\vartheta_{;\tau\sigma} \left({}^*\bar{R}^\tau_{\,\,\,\mu}{}^{\sigma}_{\,\,\,\nu} + {}^*\bar{R}^\tau_{\,\,\,\nu}{}^{\sigma}_{\,\,\,\mu}\right) \label{Geom}\\
        &&\Box \vartheta = -\frac{\alpha}{4\beta}\left[2\psi^{\mu\nu;\beta\alpha}\left({}^*\bar{R}_{\mu\alpha\nu\beta} + {}^*\bar{R}_{\mu\beta\nu\alpha}\right) \right.\nonumber\\
        &&\left.\qquad\,\,\,\,+ \bar{R}^{\alpha\beta\gamma\mu}\left({}^*\bar{R}_{\alpha}^{\,\,\,\nu}{}_{\gamma\mu} \psi_{\beta\nu} + {}^*\bar{R}_{\alpha\beta\sigma\mu}\right) \psi^\sigma_\gamma\right] \label{Deom}
        \end{eqnarray}
    \end{small}
It is important to note that, like Eqs. (\ref{eom1}), (\ref{eom2}), in the asymptotic limit, Eqs. (\ref{Geom}), (\ref{Deom}) decouple, and $ \vartheta $ is a light field.
    
Thus, the perturbed energy-momentum pseudotensor $ t_{\mu\nu} $ is of the form:
    \begin{eqnarray}
    t_{\mu\nu} &=& -\frac{\epsilon^2}{4L^2}\left<\widetilde{(\nabla\psi)^2}\right>_{\mu\nu}-\frac{\epsilon^2\kappa^2\alpha^2}{4\beta L^6} \left[\widetilde{\textbf{R}}\left<\widetilde{\nabla^2\psi \nabla^2\psi}\right>_{\mu\nu} \right.\nonumber\\
    &&\left. + \widetilde{\textbf{RR}}\left<\widetilde{\nabla\psi \nabla\psi}\right>_{\mu\nu} + \widetilde{\textbf{RRR}} \widetilde{\left<\psi\psi\right>}_{\mu\nu}\right] \nonumber\\
    &&+\frac{\epsilon^2\kappa^2\beta}{L^2}\left<\widetilde{\nabla\vartheta\nabla\vartheta}\right>_{\mu\nu} \label{effem}
    \end{eqnarray}
    where $ \psi_{\mu\nu} $, $ \vartheta $ and their derivatives have been scaled with respect to a characteristic length scale $ L $ of the background space-time (which in this case is of the order of the size of the light ring around a black-hole) such that quantities inside the angular brackets are dimensionless, and the averaging is over the short wavelength modes
    (See Appendix \ref{emp} for details). $ \widetilde{\textbf{R}} $ is the shorthand notation for dimensionless Riemann tensor. {The authors are of the opinion that (\ref{effem}) in its present form is a further simplified version of what was obtained in \cite{Stein2011a} and sheds a bit more light on on the effects of CS gravity on the EM pseudo-tensor and how the effects scale with distance from the black hole.}
    
    This is a new result of this article, regarding which we would like to stress the following points: First, the first term in the RHS of Eq.~(\ref{effem}) is the energy-momentum pseudotensor of perturbation for GR \cite{Isaacson1968b}. 
    The second and third terms in the RHS are the correction terms that arise from modification to gravity. In this case, these are the corrections from the CS gravity. 
    Second, the CS field $ \vartheta $ alone does not appear in the expression, only its derivatives. As seen from their exact forms in Appendix \ref{emp}, the contribution from the terms in the square bracket of (\ref{effem}) vanishes at a large distance from the black-hole owing to Riemann tensors appearing as a product. From bounds put in \cite{Cardoso2009,Yunes2009}, it is seen that $ \beta > \frac{\alpha^2}{\beta} $. Hence, the leading order contribution at large distances from the black-hole will come from the last term of (\ref{effem}), which is a kinetic term of the CS scalar, followed by terms with a single Riemann as a product, i.e., the first term in the square brackets of the form $ \left<\widetilde{\nabla^2\psi \nabla^2\psi}\right>_{\mu\nu} $. However, close to small black-holes, the second term can have a much stronger effect owing to the large values of curvature $ \widetilde{\textbf{R}} $ and $ L^{-6} $ dependence. The long-range effect of the last term of (\ref{effem}) is in conjunction with what was discussed at the end of the previous section.
    
    Thus a dimensionless parameter $ \Delta_{CS} $ can be defined quantifying the effect of the CS field on the gravitational perturbations near a black-hole
    \begin{eqnarray}
    \Delta_{CS} &=& \frac{\kappa^2\alpha^2}{\beta L^4} \label{delcs}
    \end{eqnarray}
{quantifying the amount to which the odd to even radiated energy ratio is suppressed in CS theories compared to GR, details of the estimation (\ref{delcs}) is given in Appendix \ref{emp}. We will discuss the relevance of the above parameter on consistence tests of GR in the next section.} 
    
    Lastly, an attentive reader would have noticed that the energy-momentum pseudotensor analysis we have obtained is independent of the background metric and holds for any black-hole space-time with a characteristic length scale ($ r_H $). In particular, the analysis applies to slowly rotating black-holes. Like in the spherically symmetric black-holes, in the slowly-rotating case, the odd and even perturbations decouple at linear order~(see, for instance, \cite{Pani:2012vp,Pani2013b,Brito2013a,Pani:2012bp}). 
    {Specifically, since the decoupling depends on the background geometry, for a slowly rotating background space-times in dynamical CS, as found in \cite{Yagi:2012ya}, the above analysis will hold as well.}

\section{Implications for future GW detectors}
    In the rest of this article, we will discuss the implications of our work for future observations to distinguish between general relativity and modified theories of gravity in general.

Using gauge-invariant formalism \cite{Martel2005} for CS gravity, we found that the parity-violating scalar field couples only to the odd parity perturbations of a spherically symmetric space-time, keeping the even parity unchanged from its GR counterpart. Similarly, the dynamics of $ \vartheta $ is also influenced by the odd parity master function, leading to the coupled system of equations (\ref{eom1}) and (\ref{eom2}). 
    
    This naturally leads to the following question: Can future detectors measure changes in energy ratio of the two opposite parity modes? To answer this question, we go back to the energy-momentum pseudo-tensor (\ref{effem}). The direct coupling of the dynamics of $ \vartheta $ with the gravitational perturbation $ \Phi_O $ allows one to write the energy-momentum tensor contribution from the CS-gravitational coupling solely in terms of the trace-reversed perturbation tensor $ \psi_{\mu\nu} $, as seen in Eq.~(\ref{effem}). Thus, RHS terms in the square bracket of (\ref{effem}) can be represented as a correction to the usual graviton-graviton interaction term in GR. The appearance of background Riemann curvatures in the stress tensor leads to the conclusion that terms appearing as a correction to GR do not propagate to asymptotic infinity. However, the effect of the CS term on gravitational radiation is strongest close to the black-hole. 
    
    A wave scattering process occurring near a black-hole will thus leak energy from the odd parity mode to the CS field. The decoupled nature of the odd and even parity modes in the linear regime will ensure no energy exchange takes place between them, {leading to an overall decrease in the net gravitational radiation and suppression of the odd to even scattered energy ratio compared to GR. The gravitational wave detectors only see the two gravitational modes, and it is possible to observe the ratio suppression directly with more detectors being planned/commissioned.}
    
    The even and odd parity modes manifest as the usual plus and cross polarizations at detectors at asymptotic infinity. The relations between the odd/even wavefunctions and the plus/cross amplitudes were given in terms of spin-weighted spherical harmonic dependence by~\cite{Nagar2005} as
    \begin{small}
    \begin{eqnarray}
    \Re\left(h_{\ell m}\right) &\simeq& \frac{1}{r}  \sqrt{\frac{\left(\ell+2\right)!}{\left(\ell-2\right)!}} \Phi_E ; \,\,\, \Im\left(h_{\ell m}\right) \simeq \frac{1}{r}  \sqrt{\frac{\left(\ell+2\right)!}{\left(\ell-2\right)!}} \Phi_O \nonumber\\
    \tilde{h}_+ - i\tilde{h}_\times &=& \sum_{\ell,m} h_{\ell m}\,\, {}_{-2}Y_{\ell m} ,
    \end{eqnarray}
    \end{small}
    where $ {}_{-2}Y_{\ell,m} $ is the spin-weighted spherical harmonic and $ \tilde{h}_{+/\times} $ are not the amplitudes of linear polarizations of GR but a circular polarization given by \cite{Jackiw2003}
    \begin{eqnarray}
    \tilde{h}_{+/\times} &=& h_{+/\times} \mp ip\dot{\vartheta} h_{\times/+} \label{circpol}
    \end{eqnarray}
    where p is the wavenumber of the pseudoscalar. We define a dimensionless parameter corresponding to the dominant multipole indices $ \ell,m $
    \begin{eqnarray}
    \Delta_{\ell,m} &=& \frac{\left|\dot{\Psi}^{\ell,m}_{O}\right|^2 - \left|\dot{\Psi}^{\ell,m}_{E}\right|^2}{\left|\dot{\Psi}^{\ell,m}_{O}\right|^2 + \left|\dot{\Psi}^{\ell,m}_{E}\right|^2} \label{dimparam}
    \end{eqnarray}
    where the overdot denotes time derivative. {From the system (\ref{eom1}), (\ref{eom2}), and the even parity dynamics, as an analogy, one can think of the net system as three harmonic oscillators, two of which are coupled (odd + CS scalar), whereas the even parity remains uncoupled from the coupled system. In GR, the net system consists of just two decoupled oscillators, for which (\ref{dimparam}) is a constant throughout the duration of the ring-down ({See Appendix \ref{tidel} for a proof}), given the dominant mode for both odd and even parities have the same multipole indices. However, due to the preferential coupling of the pseudoscalar with the odd parity in CS, $ \Delta^{CS}_{\ell, m} \leq \Delta^{GR}_{\ell, m} $, and the difference between the GR and CS values will be of the order of $ \Delta_{CS} $. Specifically, (\ref{dimparam}) will be a decreasing function of time and will be less than the corresponding GR value throughout the duration of the ring-down ({A proof of which have been given in Appendix \ref{tddel}- a feature that can act as a consistency test for GR, as well as for constraining deviations from it.}
    
    It is important to note that the factor $ip\dot{\vartheta}$ in (\ref{circpol}), imparting circular polarization to GWs at infinity do not appear in (\ref{dimparam}). More generally, the parameter $\Delta_{\ell,m}$ will not take into account equal or unequal suppression or enhancement of plus/cross polarizations from the modified theories of gravity. However, such modification to gravity do not usually arise, and they are special cases. To see this, consider the modifications from $f(R)$ and CS together. If we fine tune the coupling parameters, it is possible to have the suppression/enhancement of plus/cross with a constant $\Delta_{\ell, m}$, as in GR. However, as the reader can easily verify, these are highly unnatural. {Broadly, the most general local theory for a modification to GR (like \cite{Motohashi2018,Yunes}) will consist of parity-violating and parity non-violating sectors, whose effects on the two opposite parity massless gravitational modes may not be equal. Thus, the above parameter is a \emph{generic quantifying tool} to distinguish modified theories of gravity from GR.}  In the future gravitational wave detectors (for instance, Cosmic Explorer \cite{Evans:2016mbw}) the signal-to-noise ratio in the QNM regime could be $50$~\cite{QNMSNRbound}. These observations will help us to put a stringent bound on the factor $(\alpha^2/{\beta})$ and constrain any deviation from GR in general much better than from the currently used template matching techniques.
    
{An astute reader can make the argument that for cases like head-on collisions and radial plunges of particles into black holes, the CS pseudo-scalar won't be perturbed at all since the odd parity is not being perturbed. However, this can only be true when a non-linear regime does not precede a linear perturbation - like radial particle infall. Situations like BHBH collisions involve sufficient non-linearities, which mix opposite parities \cite{Brizuela2006,Gleiser1999}, before the system transitions to the ring-down regime. Hence for such cases, the odd parity, and consequently, the CS pseudo-scalar will always be perturbed.}
    
    The earlier analyses  \cite{Bhattacharyya2017,Bhattacharyya2018}, and the current work strongly establishes the fact that any modification to GR will lead to parity preferences of the odd and even modes and hence, {the quantity $\Delta_{\ell,m}$ will not be a constant.} The general nature of isospectrality breaking and its relations to modified theories of gravity have been discussed \cite{Barack:2018yly,Blazquez-Salcedo:2016enn}. However, to our knowledge, our work is the first to evaluate the difference and obtain a quantifying tool. 
    
    It is also possible that isospectrality breaks due to environmental contaminants around a black-hole~\cite{Barausse:2014tra}. However, environmental contaminants around a black-hole vary with each detection and would show up as different $ \Delta_{\ell,m} $ in different cases. However, modified theories of gravity will lead to a consistent non-constant value of $ \Delta_{\ell,m} $ in all observations.
    
We note again that the calculation for the energy-momentum pseudotensor is independent of the background. For slowly rotating black-holes, the odd and even metric perturbations remain decoupled and hence, our current analysis holds. However, the same is not evident for fast rotating space-times where the Pontryagin density does not vanish in the background, and the authors are not aware of any non-perturbative (in spin) axisymmetric black hole solutions in dynamical CS gravity. Although, a feature that's noticeable while perturbing space-times of generic spins in GR is that gravitational perturbations of opposite parities and the same multipoles do not mix \cite{Pani:2012bp}- indicating that the dominant odd parity $ \left(\ell,m\right) = \left(2,2\right) $ mode does not mix with the even parity mode of the same multipolar indices. One possible future work will be to obtain the $ \left(2,2\right) $ calculation of (\ref{dimparam}) using numerical relativity data of final states of binary black-hole mergers and precisely evaluate the changes between GR and modified theories of gravity.  Simulations of the CS system, like the one proposed in \cite{Okounkova:2018pql}, will also help us understand the dynamics and asymptotic behavior of the radiation associated with the CS field, i.e., the last term of (\ref{effem}). Detectors with better sensitivity towards scalar degrees of freedom can then technically probe for scalar radiation from scattering processes or merger events. {Since the CS pseudoscalar is a massless field, its kinetic term will have a markedly stronger effect at long ranges compared to the coupling terms} and hence can help us ascertain whether scalar fields (such as the CS field) exists in the Universe. 
    
\begin{acknowledgments}
  We have used Cadabra~\cite{Peeters:2007wn,PEETERS2007550} for the calculations. 
        The authors would like to thank Akihiro Ishibashi, Bala Iyer, and Badri Krishnan for clarifications. SB is financially supported by the MHRD fellowship at IISER-TVM and would like to thank P Ajith's group at ICTS for hospitality. SS thanks University of Lethbridge and Shastri Indo-Canadian Institute for the hospitality during the writing of this paper. SS is partially supported by Homi Bhabha Fellowship. 
\end{acknowledgments}
\appendix
\section{Various first order perturbed 2+2 decomposed quantities}
\subsection{Perturbed Pontryagin density}\label{ptbpont}
The perturbed metric tensor can be 2+2 decomposed following \cite{Kodama:2000fa}
\begin{eqnarray}
h_{\mu\nu} \equiv \left(\begin{array}{cc}
f_{ab} \textbf{S}& f_a^E\textbf{S}_A + f_a^O \textbf{V}_A \\ 
Sym & H_T^E \textbf{S}_{AB} + H_T^O \textbf{V}_{AB} + H_L^E \gamma_{AB} \textbf{S}
\end{array}\right)\nonumber\\ \label{metdecomp}
\end{eqnarray}
where $ f_{ab} $, $ f_a^E $, $ f_a^O $, $ H_T^E $, $ H_T^O $, and $ H_L^E $ are a set of ten scalars only dependent on $ \left(t,r\right) $ (subscript T and L imply transverse and longitudinal components respectively). An implicit summation over $ \ell $, $ m $ was assumed in (\ref{metdecomp}). $ \textbf{S} $, $ \textbf{S}_A $, and $ \textbf{S}_{AB} $ are even parity spherical harmonic scalar, vector, and tensor respectively. $ \textbf{V}_A $ and $ \textbf{V}_{AB} $ are odd parity spherical harmonic vector and tensor respectively. The odd parity spherical harmonic vector on a 2-sphere is related to the even parity spherical harmonic scalar as
\begin{eqnarray}
\textbf{V}_A &=& \epsilon_{AB} \hat{D}^B \textbf{S} = \epsilon_{AB} \partial^{B} \textbf{S} \label{oerel}
\end{eqnarray}
as defined in \cite{Martel2005}, where $ \epsilon_{AB} $ and $ \hat{D} $ are the the covariant Levi-Civita density and the covariant derivative defined on a 2-sphere respectively.

A covariant Levi-Civita on a 2-sphere can be constructed by projecting out of a Levi-Civita in the full space-time as
\begin{eqnarray}
\epsilon_{AB} &=& \frac{1}{\sqrt{2}r^2} \epsilon_{AaBb} \epsilon^{ab} \label{2slc}
\end{eqnarray}
where $ \epsilon_{ab} $ is the covariant Levi-Civita on the $ \left(t,r\right) $ space. (\ref{2slc}) satisfies all the properties of the antisymmetric 2-form in the 2-sphere. Using (\ref{metdecomp}), (\ref{oerel}), and (\ref{2slc}), the perturbed Pontryagin density for a background Schwarzschild space-time in terms of the \emph{Cunningham-Price-Moncrief} variable $ \Phi_O $ (as defined in \cite{Martel2005}) becomes
\begin{eqnarray}
\delta\left({}^*RR\right) &=& \frac{24 \left(\ell-1\right) \ell \left(\ell+1\right) \left(\ell+2\right) M}{r^6} \,\, \Phi_O \,\, \textbf{S}
\end{eqnarray}
where $ S_{\ell m} $ is the scalar spherical harmonic and it's seen that only the odd parity master function contributes to the perturbed Pontryagin density.
\subsection{Perturbed Cotton tensor as an effective source.}\label{ptbcotton}
The perturbed Cotton tensor can be written as an effective energy-momentum tensor in the following manner
\begin{eqnarray}
R^{(1)}_{\mu\nu} &=& \kappa^2 \mathcal{T}_{\mu\nu} \\
\mathcal{T}_{\mu\nu} &=& -\alpha \Theta_{;\tau\sigma}\left( {}^*\bar{R}^\tau_{\,\,\,\mu\nu}{}^\sigma + {}^*\bar{R}^\tau_{\,\,\,\nu\mu}{}^\sigma \right)
\end{eqnarray}
A vector and a scalar can be defined from $ T^{eff}_{\mu\nu} $ in the following manner, following \cite{Martel2005}
\begin{eqnarray}
P^a &=& \frac{\kappa^2r^2}{k^2} \int \mathcal{T}^{a A} V_{A} \,\, d\Omega \\
P &=&\frac{\kappa^2r^4}{\left(\ell-1\right)\ell\left(\ell+1\right)\left(\ell+2\right)} \int \mathcal{T}^{AB} V_{AB} \,\, d\Omega
\end{eqnarray}
Using $ \Theta = \frac{\psi}{r} \textbf{S} $, we obtain the following
\begin{eqnarray}
P^t &=& -\frac{6M}{r^2} \partial_r \varphi \\
P^r &=& \frac{6i\omega M}{r^2} \varphi \\
P &=& 0
\end{eqnarray}
The above components satisfy the conservation equation $ \nabla^\mu \mathcal{T}_{\mu\nu} = 0 $. Following \cite{Martel2005}, we find
\begin{eqnarray}
\partial_t P^t + \partial_r P^r + \frac{2}{r} P^r &=& 0
\end{eqnarray}
which serves as a consistency check for the obtained components. Again, following \cite{Martel2005}, the effective source term coupling with the odd parity gravitational perturbation was found to be
\begin{eqnarray}
S^{eff} &=& \frac{\kappa^2\alpha}{\left(\ell-1\right)\left(\ell+2\right)} \left[\frac{6M}{r}\partial^2_{r_*}\varphi - \frac{12M}{r^2} \partial_{r_*}\varphi + \frac{6\omega^2M}{r}\varphi\right]\nonumber\\
\end{eqnarray}
\section{Gravitational radiation in the shortwave limit}\label{emp}
A vanishing background $ \vartheta $ and transverse-traceless gauge  was used. For a metric and CS field perturbation
\begin{eqnarray}
g_{\mu\nu} &=& \bar{g}_{\mu\nu} + e h_{\mu\nu} \\
\vartheta &=& e \vartheta,
\end{eqnarray}
the modified field tensor $ \mathfrak{G}_{\mu\nu} = R_{\mu\nu} 2\kappa^2\alpha C_{\mu\nu} - \kappa^2\beta\vartheta_{;\mu}\vartheta_{;\nu} $ can be expanded in powers of $ e $ as
\begin{eqnarray}
\mathfrak{\bar{G}}_{\mu\nu} + e \mathfrak{G}^{(1)}_{\mu\nu} + e^2 \mathfrak{G}^{(2)}_{\mu\nu} &=& 0 \label{eseries}
\end{eqnarray}
Solving for $ \mathfrak{G}^{(1)}_{\mu\nu} = 0 $ gives the dynamics of the perturbation. While the radiated energy and momentum flux due to perturbation can be found from an energy-momentum pseudotensor due to perturbation. From (\ref{eseries}) we then get
\begin{eqnarray}
\bar{\mathfrak{G}}_{\mu\nu} &=& \kappa^2t_{\mu\nu} \\
&=& -e^2\left<\mathfrak{G}^{(2)}_{\mu\nu}\right> \\
t_{\mu\nu} &=& -\frac{e^2}{\kappa^2} \left<\mathfrak{G}^{(2)}_{\mu\nu}\right> \\
\mathfrak{G}^{(2)}_{\mu\nu} &=& G^{(2)}_{\mu\nu} - 4\kappa^2\alpha \left[\nabla^{(1)}_\sigma\nabla_\tau\vartheta\,\,{}^*\bar{R}^\tau_{\,\,\,\left(\mu\right.}{}^{\sigma}_{\,\,\,\left.\nu\right)}\right.\nonumber\\
&&\left. + \nabla_\sigma\nabla_\tau\vartheta\,\,{}^*R^{(1)\tau}_{\quad\,\,\left(\mu\right.}{}^\sigma_{\,\,\,\left.\nu\right)}\right] - \kappa^2\beta\vartheta_{;\mu}\vartheta_{;\nu}
\end{eqnarray}
$ \left<...\right> $ was defined in \cite{Isaacson1968a} and consists of the following effective operations
\begin{itemize}
	\item Total derivative terms are put to zero.
	\item $ \left<A_{;\mu}B_{;\nu}\right> = -\left<A_{;\mu\nu}B\right>  $
	\item Covariant derivatives commute.
	\item Average of a product of two different fields are put to zero, since for high frequencies they are Gaussian random variables.
\end{itemize}
From which $ \left<\mathfrak{G}^{(2)}_{\mu\nu}\right> $ was found to be
\begin{eqnarray}
-\left<\mathfrak{G}^{(2)}_{\mu\nu}\right> &=& \frac{1}{4}\left<\psi^{\rho\tau}_{;\mu}\psi_{\rho\tau;\nu}\right> + \frac{\kappa^2\alpha^2}{2\beta}\left<\mathcal{P}_{\mu\nu}\right> - \kappa^2\beta\left<\vartheta_{;\mu}\vartheta_{;\nu}\right>\nonumber\\
\\
\left<\mathcal{P}_{\mu\nu}\right> &=& -2 \left<\psi^{\beta\gamma;\delta\lambda} \psi_{\nu\alpha;\sigma}{}^{;\rho}\right> \epsilon_{\mu\rho}{}^{\sigma\alpha}\left({}^*\bar{R}_{\lambda\beta\delta\gamma} + {}^*\bar{R}_{\delta\beta\lambda\gamma}\right) \nonumber\\
&&-2\left<\psi^{\beta\gamma;\delta\lambda}\psi_{\mu\alpha;\sigma}{}^{;\rho}\right>\epsilon_{\nu\rho}{}^{\sigma\alpha} \left({}^*\bar{R}_{\lambda\beta\delta\gamma} + {}^*\bar{R}_{\delta\beta\lambda\gamma}\right)\nonumber\\
&&-2\left<\psi^{\rho\sigma;\delta}\psi^{\alpha\beta;\gamma}\right>\left[{}^*\bar{R}_{\gamma\alpha\delta\beta}\left({}^*\bar{R}_{\mu\sigma\nu\rho} + {}^*\bar{R}_{\nu\sigma\mu\rho}\right)\right.\nonumber\\
&&\left. + {}^*\bar{R}_{\delta\alpha\gamma\beta}\left({}^*\bar{R}_{\mu\sigma\nu\rho} + {}^*\bar{R}_{\nu\sigma\mu\rho}\right)\right] \nonumber \\
&&+\bar{R}^{\rho\sigma\alpha\beta}\left[\epsilon_{\mu\gamma}{}^{\delta\lambda}\left(\left<\psi^\eta_\sigma{}^{;\gamma}\psi_{\nu\lambda;\delta}\right>{}^*\bar{R}_{\rho\eta\alpha\beta}\right.\right.\nonumber\\
&&\left.+ \left<\psi^\eta_\alpha{}^{;\gamma}\psi_{\nu\lambda;\delta}\right>{}^*\bar{R}_{\rho\sigma\eta\beta}\right)\nonumber\\
&&+ \epsilon_{\nu\lambda}{}^{\delta\gamma}\left(\left<\psi^\eta_\sigma{}^{;\gamma}\psi_{\mu\lambda;\delta}\right>{}^*\bar{R}_{\rho\eta\alpha\beta}\right.\nonumber\\
&&\left.\left. + \left<\psi^\eta_\alpha{}^{;\gamma}\psi_{\mu\lambda;\delta}\right>{}^*\bar{R}_{\rho\sigma\eta\beta}\right)\right]\nonumber\\
&&+ \bar{R}^{\rho\sigma\alpha\beta}\left[\left<\psi^{\gamma\delta}\psi^\lambda_\alpha\right>{}^*\bar{R}_{\rho\sigma\lambda\beta}\left({}^*\bar{R}_{\mu\delta\nu\gamma} + {}^*\bar{R}_{\nu\delta\mu\gamma}\right)\right.\nonumber\\
&&\left.+\left<\psi^{\gamma\delta}\psi^\lambda_\sigma\right>{}^*\bar{R}_{\rho\lambda\alpha\beta}\left({}^*\bar{R}_{\mu\delta\nu\gamma} + {}^*\bar{R}_{\nu\delta\mu\gamma}\right)\right] \label{csem}
\end{eqnarray}
An estimate for the leading order power density for the coupling term can be obtained from (\ref{csem}) using the 00 component. If we consider a background space-time with characteristic length scale $ L $, the metric perturbation and the background Riemann tensor will be of the form
\begin{eqnarray}
\psi_{\mu\nu} &\sim& \frac{L}{r} \\
\bar{R}_{\mu\nu\rho\sigma} &\sim& \frac{L}{r^3}
\end{eqnarray}
such that the leading order term in $ \left<\mathcal{P}_{00}\right> $ become
\begin{eqnarray}
\left<\mathcal{P}_{00}\right> &\sim& \frac{\kappa^2\alpha^2}{\beta} \frac{1}{Lr^5}
\end{eqnarray}
and the first (GR) term of (\ref{csem}) can be represented as
\begin{eqnarray}
\left<\psi^{\rho\tau}_{;0}\psi_{\rho\tau;0}\right> &\sim& \frac{1}{r^2}
\end{eqnarray}
from where we take the ratio of the two terms and define
\begin{eqnarray}
\Delta_{CS} &=& \frac{\kappa^2\alpha^2}{\beta} \frac{1}{Lr^3}
\end{eqnarray}
Scaling the radial variable with respect to the background characteristic length scale as $ r = y L $,
\begin{eqnarray}
\Delta_{CS} &=& \frac{\kappa^2\alpha^2}{\beta L^4} \frac{1}{y^3}
\end{eqnarray}
the $ \frac{1}{y^3} $ dimensionless factor can be integrated out to give a factor which won't change the approximate order of $ \Delta_{CS} $, hence we obtain,
\begin{eqnarray}
\Delta_{CS} &=& \frac{\kappa^2\alpha^2}{\beta L^4}
\end{eqnarray}
\section{Matching with plus and cross at asymptotic infinity} \label{oepc}
In order to equate the parity polarizations with the plus and cross, we project the radiative part of the metric perturbation on a tetrad of freely falling observers in the radiation zone. The radiative part is simply the perturbation about the background 2-sphere in the 2+2 decomposed metric. We have from \cite{Martel2005,Nagar2005,Jackiw2003,Alexander:2009tp}
\begin{eqnarray}
h_{\hat{A}\hat{B}} &=& e^A_{\hat{A}} e^B_{\hat{B}} h_{AB} \\
&=& \frac{\Phi_E}{r} \left(\begin{array}{cc}
\textbf{S}_{\theta\theta} & \frac{\textbf{S}_{\theta\phi}}{\sin\theta} \\ 
\frac{\textbf{S}_{\theta\phi}}{\sin\theta} & \frac{\textbf{S}_{\phi\phi}}{\sin^2\theta}
\end{array} \right)  + \frac{\Phi_O}{r} \left(\begin{array}{cc}
\textbf{V}_{\theta\theta} & \frac{\textbf{V}_{\theta\phi}}{\sin\theta}  \\ 
\frac{\textbf{V}_{\theta\phi}}{\sin\theta} & \frac{\textbf{V}_{\phi\phi}}{\sin^2\theta}
\end{array} \right)\nonumber\\ \label{paras}\\
&=&\left(\begin{array}{cc}
h_+ - ip\dot{\vartheta}h_\times & h_\times + ip\dot{\vartheta}h_+ \\ 
h_\times + ip\dot{\vartheta}h_+ & -h_+ + ip\dot{\vartheta}h_\times
\end{array} \right) \label{pc} \\
&=&\left(\begin{array}{cc}
\tilde{h}_+ & \tilde{h}_\times \\ 
\tilde{h}_\times & -\tilde{h}_+ \label{ppc}
\end{array} \right)
\end{eqnarray}
where an implicit summation of $ \ell, m $ was assumed and $ p $ is the wavenumber corresponding the plus/cross polarizations. Comparing (\ref{paras}) and (\ref{ppc}) and using the relation between the tensor spherical harmonics and spin-weighted spherical harmonics \cite{Nagar2005} obtains
\begin{eqnarray}
\tilde{h}_+ - i\tilde{h}_\times &\simeq& \frac{1}{r} \sum_{\ell,m} \sqrt{\frac{\left(\ell+2\right)!}{\left(\ell-2\right)!}} \left(\Phi_E + i\Phi_O\right)\,{}_{-2}Y_{\ell m} \nonumber\\\label{n1}
\end{eqnarray}
LHS is the doubly integrated Weyl scalar $ \Psi_4 $ at asymptotic infinity, and can be expanded in spin-weighted spherical harmonics as
\begin{eqnarray}
h_+ - ih_\times &=& \sum_{\ell,m} h_{\ell m} \,{}_{-2}Y_{\ell m}. \label{o1}
\end{eqnarray}
Comparing (\ref{n1}) and (\ref{o1}) we have,
\begin{eqnarray}
\Re\left(h_{\ell m}\right) &\simeq& \frac{1}{r} \sum_{\ell,m} \sqrt{\frac{\left(\ell+2\right)!}{\left(\ell-2\right)!}} \Phi_E \\
\Im\left(h_{\ell m}\right) &\simeq& \frac{1}{r} \sum_{\ell,m} \sqrt{\frac{\left(\ell+2\right)!}{\left(\ell-2\right)!}} \Phi_O
\end{eqnarray}
\section{Constancy of $ \Delta_{\ell m} $ in GR and its time dependence in CS}
\subsection{Constancy in GR}\label{tidel}
The quantity $ \Delta_{\ell m} $ in the main text given by
\begin{eqnarray}
\Delta_{\ell,m} &=& \frac{\left|\dot{\Psi}^{\ell,m}_{O}\right|^2 - \left|\dot{\Psi}^{\ell,m}_{E}\right|^2}{\left|\dot{\Psi}^{\ell,m}_{O}\right|^2 + \left|\dot{\Psi}^{\ell,m}_{E}\right|^2} 
\end{eqnarray}
can be written as
\begin{eqnarray}
\Delta_{\ell,m} &=& \frac{\frac{\left|\dot{\Psi}^{\ell,m}_{O}\right|^2}{\left|\dot{\Psi}^{\ell,m}_{E}\right|^2} - 1}{\frac{\left|\dot{\Psi}^{\ell,m}_{O}\right|^2}{\left|\dot{\Psi}^{\ell,m}_{E}\right|^2} + 1} \label{dimparam1}
\end{eqnarray}
In the wave zone, the odd/even modes are of the form
\begin{eqnarray}
\Psi_{E/O} &=& A_{E/O} e^{-\kappa_{E/O}t} e^{i\omega_{E/O}t}  \label{asympgr}
\end{eqnarray}
where $ A_{E/O} $ is a constant amplitude that depends on the initial conditions of the perturbation process. Due to isospectrality relation for GR, $ \kappa_{E} = \kappa_O =\kappa $ and $ \omega_E=\omega_O = \omega $. Substituting (\ref{asympgr}) in (\ref{dimparam1}) one obtains
\begin{eqnarray}
\Delta_{\ell,m} &=& \frac{\left|\frac{A_O}{A_E}\right|^2-1}{\left|\frac{A_O}{A_E}\right|^2+1}
\end{eqnarray}
which is a constant.
\subsection{Time dependent $ \Delta_{\ell,m} $ in CS gravity}\label{tddel}
Radiation rate escaping to asymptotic infinity for general relativity is given by \cite{Martel2005}
\begin{eqnarray}
\left.\left<\dot{E}\right>\right|_{GR} &=& \frac{1}{64\pi} \sum_{\ell m} \mu\left<\left|\dot{\Psi}_E\right|^2 + \left|\dot{\Psi}_O\right|^2\right> \\
\mu &=& \left(\ell-1\right)\ell\left(\ell+1\right)\left(\ell+2\right)
\end{eqnarray}
Similarly, for dynamical CS gravity the rate at which radiation (both gravitational and scalar) escapes to asymptotic infinity can be given by
\begin{eqnarray}
\left.\left<\dot{E}\right>\right|_{CS} &=& \frac{1}{64\pi} \sum_{\ell m} \mu\left<\left|\dot{\Psi}_E\right|^2 + \left|\dot{\tilde{\Psi}}_O\right|^2 + \kappa^2\beta \left|\dot{\varphi}\right|^2\right> \nonumber\\
\end{eqnarray}
There is also energy loss $ \left<\dot{E}_{coup,CS}\right> $ in the form of the graviton-graviton coupling near the BH region (\ref{csem}) which does not travel to asymptotic infinity, thereby effectively reducing the odd parity reflection coefficient, or the fraction of the odd parity initial excitation that gets scattered off to asymptotic infinity, compared to GR. Considering the same initial perturbation energy for a Schwarzschild solution in GR and dynamical CS, the latter shall then radiate lesser gravitational flux, with the difference in energy coming from both the graviton-graviton coupling (which absorbed by the BH), and the kinetic term of the pseudoscalar field. Thus we can write the following
\begin{eqnarray}
\left.\left<\dot{E}\right>\right|_{CS} + \left.\left<\dot{E}\right>\right|_{coup,CS} &=& \left.\left<\dot{E}\right>\right|_{GR}
\end{eqnarray}
from which one obtains the following inequality
\begin{eqnarray}\label{ineq}
\left|\dot{\Psi}_O\right|^2 > \left|\dot{\tilde{\Psi}}_O\right|^2
\end{eqnarray}
at all times. A suitable ansatz for the modified odd parity wavefunction for CS gravity can be
\begin{eqnarray}
\tilde{\Psi}_O &=& \tilde{A}_Oe^{-\tilde{\kappa}_Ot}e^{i\tilde{\omega}_Ot}
\end{eqnarray}
where $ \tilde{A}_O < A_O $, the real and imaginary parts of the odd parity QNM frequency are modified due to the coupling with the CS field in the form of an inhomogeneous term in the RHS of the differential equation (9) in the main text. This leads to the following
\begin{eqnarray}
\frac{\left|\dot{\tilde{\Psi}}_O\right|^2}{\left|\dot{\Psi}_E\right|^2} &=& \frac{\tilde{A}_O^2\left(\tilde{\kappa}_O^2 + \tilde{\omega}_O^2\right)}{A_E^2\left(\kappa_E^2 + \omega_E^2\right)} e^{-2\left(\tilde{\kappa}_O - \kappa_E\right)t} \label{model}
\end{eqnarray}
which is less than the corresponding GR value at all times courtesy (\ref{ineq}), with a growth/decay rate proportional to $ e^{-2\left(\tilde{\kappa}_O - \kappa_E\right)t} $ (depending on whether the imaginary part of the odd parity dominant mode frequency is enhanced or suppressed due to CS modification). However, for the same initial energy of perturbation, the odd parity mode can now relax to a stable Schwarzschild faster, because of the presence of further channels (pseudo-scalar and graviton-graviton coupling) to take away the initial perturbation energy. This leads to a shorter modified decay time for the odd parity mode compared to the even parity~\footnote{A feature that is also seen in charged BHs in GR, with the damping time decreasing with increase in charge (See Table I of \cite{Leaver:1990zz}). This is because, for a purely gravitational perturbation, extra energy in the system can now be radiated away through the electromagnetic waves as well (in addition to gravitational degrees of freedom), making the relaxation to a stable solution faster. Although the signatures imparted to gravitational waves due to the presence of charge is quite distinct from the signatures imparted due to modifications to gravity. Whereas isospectrality holds for perturbed charged GR BHs, the same does not hold for a perturbed Schwarzschild in CS modified gravity.}, i.e. $ \tilde{\kappa}_O > \kappa_{E} $ --- implying that (\ref{model}), and correspondingly $ \Delta_{\ell,m} $, will be decreasing functions of time in CS gravity.
%

\end{document}